\documentclass[a4paper]{jpconf}
\usepackage{graphicx}
\usepackage{xspace}
\usepackage{amsmath}
\usepackage{amssymb}

\newcommand{\fbinv} {\mbox{\ensuremath{\,\text{fb}^{-1}}}\xspace}
\newcommand{\GeV}{\ensuremath{\,\mathrm{Ge\hspace{-.08em}V}}\xspace}
\newcommand{\Pt}{\ensuremath{p_{\mathrm{T}}}\xspace}
\newcommand{\ttbar}{t\ensuremath{\mathrm{\bar{t}}}\xspace}
\newcommand{\mtplx}{\textsc{Matriplex}}
\newcommand{\mkFit}{\textsc{mkFit}}

\begin{document}

\title{Speeding up the CMS track reconstruction with a parallelized and
vectorized Kalman-filter-based algorithm during the LHC Run 3}

\author{S Berkman, G Cerati$^{\,1}$, P Elmer$^{\,2}$, P Gartung$^{\,1}$, L Giannini$^{\,3}$, B Gravelle$^{\,4}$, A R Hall$^{\,5}$, M Kortelainen$^{\,1}$, S Krutelyov$^{\,3}$, S R Lantz$^{\,6}$, M Masciovecchio$^{\,3}$, K McDermott, B Norris, M Reid$^{\,6}$, D S Riley$^{\,6}$, M Tadel$^{\,3}$, E Vourliotis$^{\,3a}$, B Wang, P Wittich$^{\,6}$, A Yagil$^{\,3}$\\
on behalf of the CMS Collaboration}

\address{$^{1\,}$Fermilab, IL, US\\
$^{2\,}$Princeton University, NJ, US\\
$^{3\,}$University of California San Diego, CA, US\\
$^{4\,}$University of Oregon, OR, US\\
$^{5\,}$USNA Annapolis, MD, US\\
$^{6\,}$Cornell University, NY, US}

\ead{$^{a\,}$emmanouil.vourliotis@cern.ch}

\begin{abstract}
One of the most challenging computational problems in the Run 3 of the Large Hadron Collider (LHC) and more so in the High-Luminosity LHC (HL-LHC) is expected to be finding and fitting charged-particle tracks during event reconstruction. The methods used so far at the LHC and in particular at the CMS experiment are based on the Kalman filter technique. Such methods have shown to be robust and to provide good physics performance, both in the trigger and offline. In order to improve computational performance, we explored Kalman-filter-based methods for track finding and fitting, adapted for many-core SIMD architectures. This adapted Kalman-filter-based software, called ``\mkFit", was shown to provide a significant speedup compared to the traditional algorithm, thanks to its parallelized and vectorized implementation. The \mkFit\ software was recently integrated into the offline CMS software framework, in view of its exploitation during the Run 3 of the LHC. At the start of the LHC Run 3, \mkFit\ will be used for track finding in a subset of the CMS offline track reconstruction iterations, allowing for significant improvements over the existing framework in terms of computational performance, while retaining comparable physics performance. The performance of the CMS track reconstruction using \mkFit\ at the start of the LHC Run 3 is presented, together with prospects of further improvement in the upcoming years of data taking.
\end{abstract}

\section{Motivation and the \mkFit\ Algorithm} \label{sec:MotivationAndMkFit}

As the Large Hadron Collider (LHC) experiments prepare for the Phase-2 upgrade of the particle accelerator with the ultimate goal of gathering about $3000 \fbinv$ of integrated luminosity, studies are performed to estimate the computational resources needed for the reconstruction of collision events. These studies indicate a superlinear growth for the total reconstruction time~\cite{HighPUTracking}. Among the different aspects of the reconstruction process, the reconstruction of charged particle trajectories, simply called ``tracks", takes almost half of the total time of the current Run-3 reconstruction~\cite{mkFitDP}. To complement single-threaded runtime performance, it is clear parallelized and vectorized tracking algorithms need to be developed. These will be crucial for Phase-2 but they can make a difference even in Run-3 of the LHC.

\begin{figure}[tbh!]
    \begin{center}
        \includegraphics[width=0.60\textwidth]{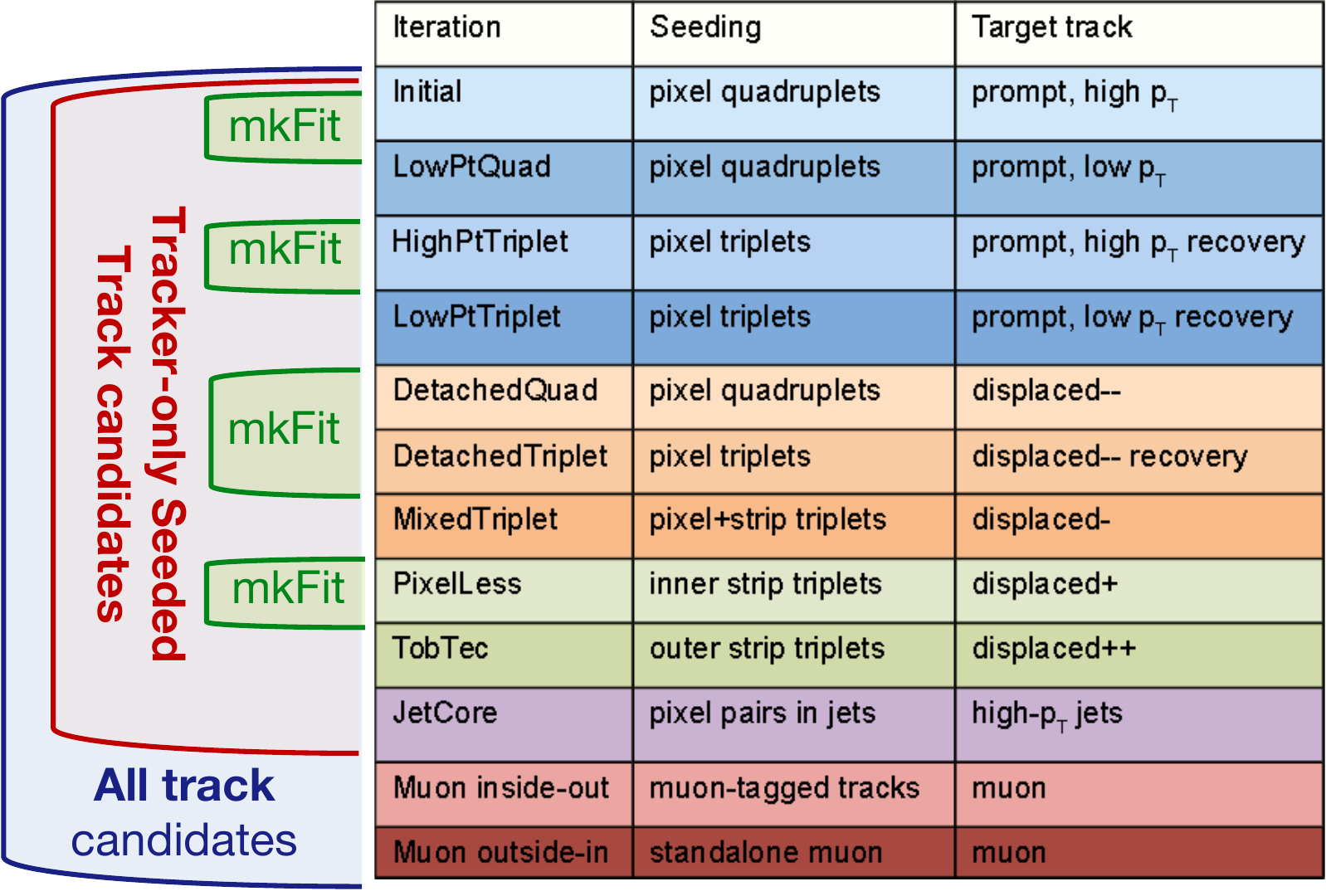}
        \hspace{2pc}
        \begin{minipage}[b]{0.33\textwidth}
            \caption{\label{fig:mkFitTrackingIterations}Iterations of the CMS track reconstruction sequence. The first column indicates the name of the iteration, the second column the kind of tracks used for the seeding of the full track reconstruction and the third column the track type that each iteration targets. On the left, the iterations for which the \mkFit\ algorithm is used are marked.}
        \end{minipage}
    \end{center}
\end{figure}

The \mtplx\ Kalman-fitter algorithm, ``\mkFit" for short, is a parallelized and vectorized version of the combinatorial Kalman-filter (CKF) algorithm used for the track reconstruction at the Compact Muon Solenoid (CMS) experiment at the LHC~\cite{CMSExperiment,CMSTrackingPaper}. Being in development for more than five years, it has been integrated during the first quarter of 2022 in the production release of CMS and used for the entire first year of Run 3 data taking. It has been applied to a subset of the CMS track reconstruction iterations, as illustrated in figure~\ref{fig:mkFitTrackingIterations}, and achieves similar physics performance as the CKF algorithm, while providing a significant speed up, as shown in section~\ref{sec:PhysicsTimingPerformance}.

In a nutshell, the CKF algorithm starts from track seeds and iteratively accumulates compatible hits to build full tracks. The \mkFit\ algorithm speeds up this procedure by vectorizing some of its aspects and making them multithreaded. For this to happen, some requirements need to be fulfilled: the branching points of the algorithm need to be minimized, the workload needs to be equally distributed to different threads and the memory accesses need to be minimized and optimized. To reduce the branching points in the code, the \mkFit\ algorithm cleverly parallelizes the track building over multiple levels (different events, different $\eta$ regions and different $z$-/$r$- and $\phi$-sorted groups of seeds). The balancing of the workloads is achieved with the usage of the Intel$^\circledR$ Threading Building Blocks (TBB) library. In terms of memory usage, this is greatly improved by utilizing a custom matrix library, \mtplx, specially designed to optimize the memory accesses for the $6 \times 6$ track candidate covariance matrices used during Kalman filter operations~\cite{mkFitPaper}. Finally, the memory needs are vastly reduced by dropping the detailed information on individual tracker modules in favor of a simplified tracker geometry, in which the tracker details are stored in a 2D $(r~\text{or}~z,\phi)$ map.

\section{Physics and Timing Performance} \label{sec:PhysicsTimingPerformance}

\begin{figure}[tbh!]
\begin{center}
    \begin{minipage}{0.45\textwidth}
        \includegraphics[width=0.95\textwidth]{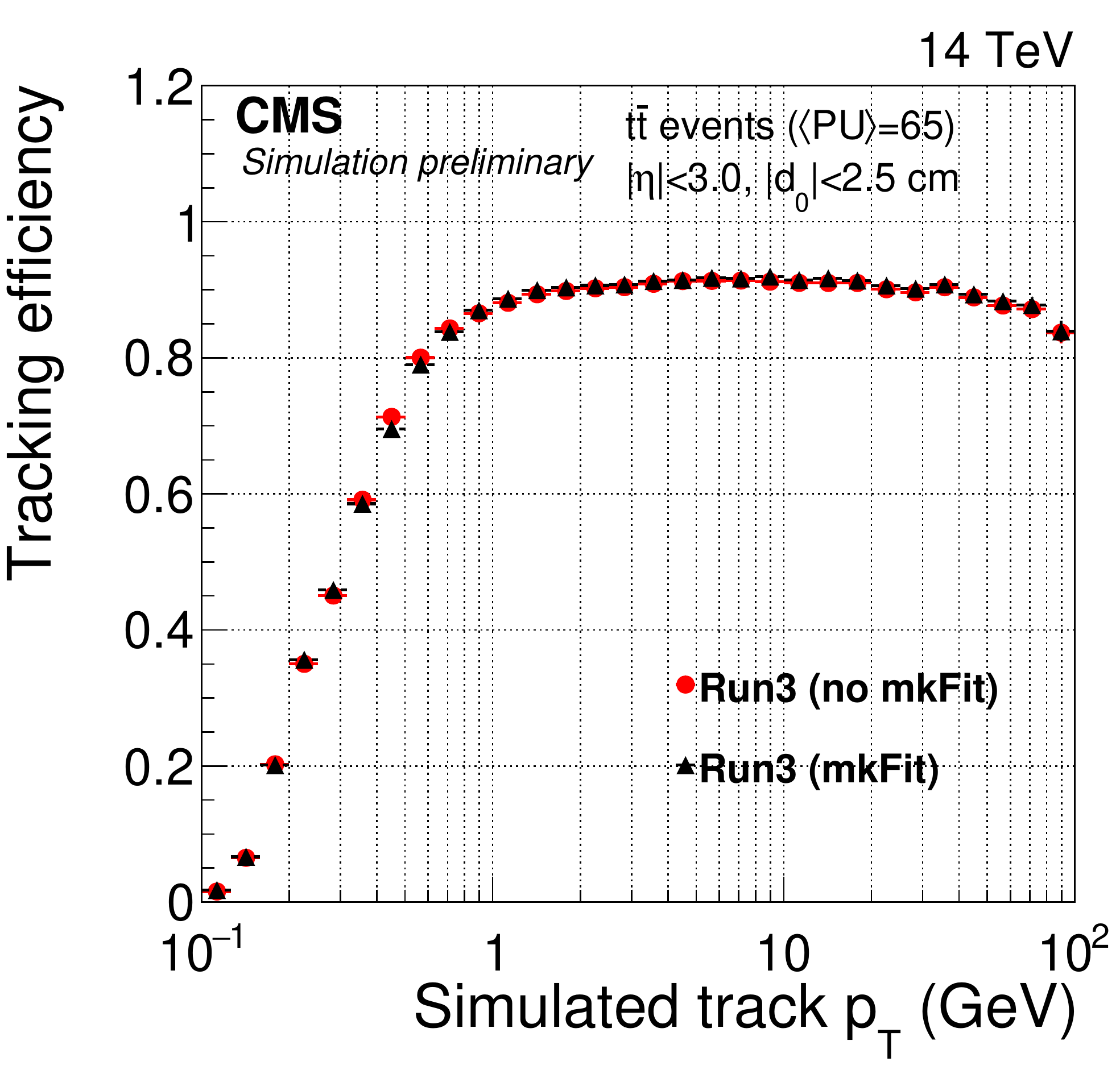}
        \vspace{-1ex}
        \caption{\label{fig:efficiency_pt}Tracking efficiency as a function of the simulated track \Pt for CKF tracking (red) and \mkFit\ tracking (black), for simulated tracks with $|{\eta}| < 3$ and $|d_{0}| < 2.5~\text{cm}$~\cite{mkFitDP}.}
    \end{minipage}
    \hspace{2pc}
    \begin{minipage}{0.43\textwidth}
        \includegraphics[width=\textwidth]{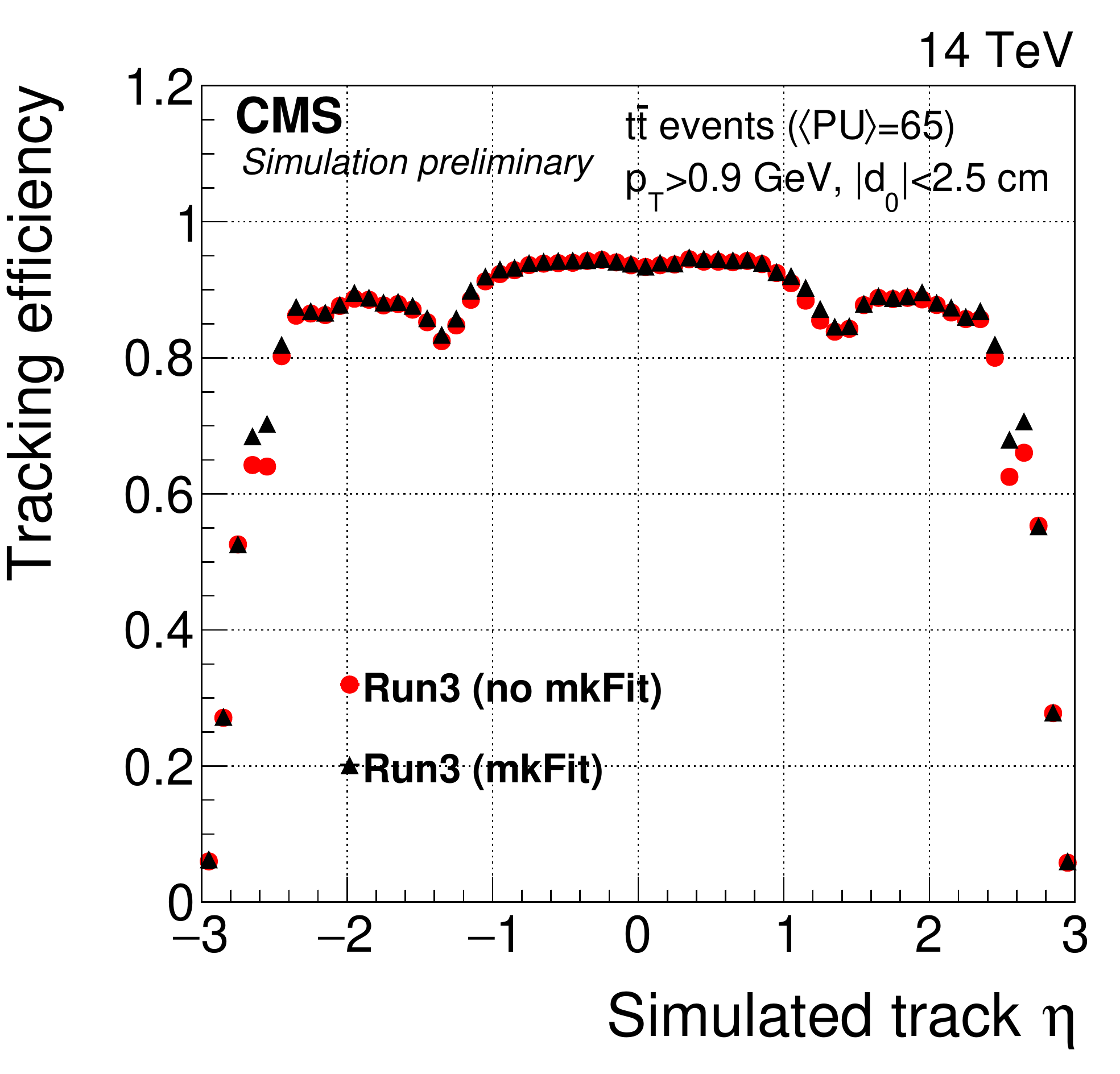}
        \vspace{-4ex}
        \caption{\label{fig:efficiency_eta}Tracking efficiency as a function of the simulated track $\eta$ for CKF tracking (red) and \mkFit\ tracking (black), for simulated tracks with $\Pt > 0.9 \GeV$ and $|d_{0}| < 2.5~\text{cm}$~\cite{mkFitDP}.}
    \end{minipage}
    \vspace{-5ex}
\end{center}
\end{figure}

\begin{figure}[tbh!]
\begin{center}
    \begin{minipage}{0.45\textwidth}
        \includegraphics[width=0.95\textwidth]{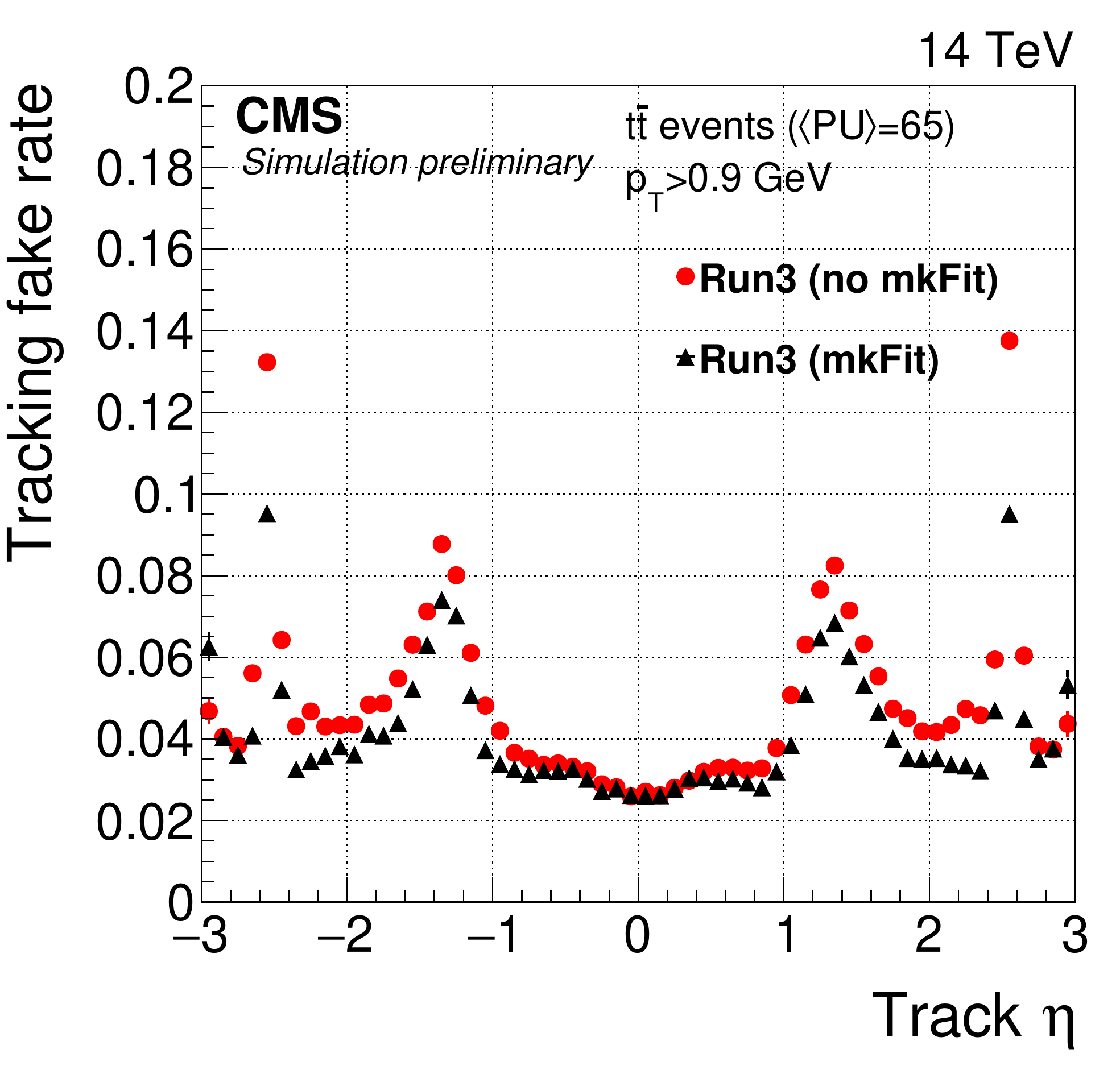}
        \vspace{-1ex}
        \caption{\label{fig:fakerate_eta}Tracking fake rate as a function of the reconstructed track $\eta$ for CKF tracking (red) and \mkFit\ tracking (black), for reconstructed tracks with $\Pt > 0.9 \GeV$~\cite{mkFitDP}.}
    \end{minipage}
    \hspace{2pc}
    \begin{minipage}{0.43\textwidth}
        \includegraphics[width=\textwidth]{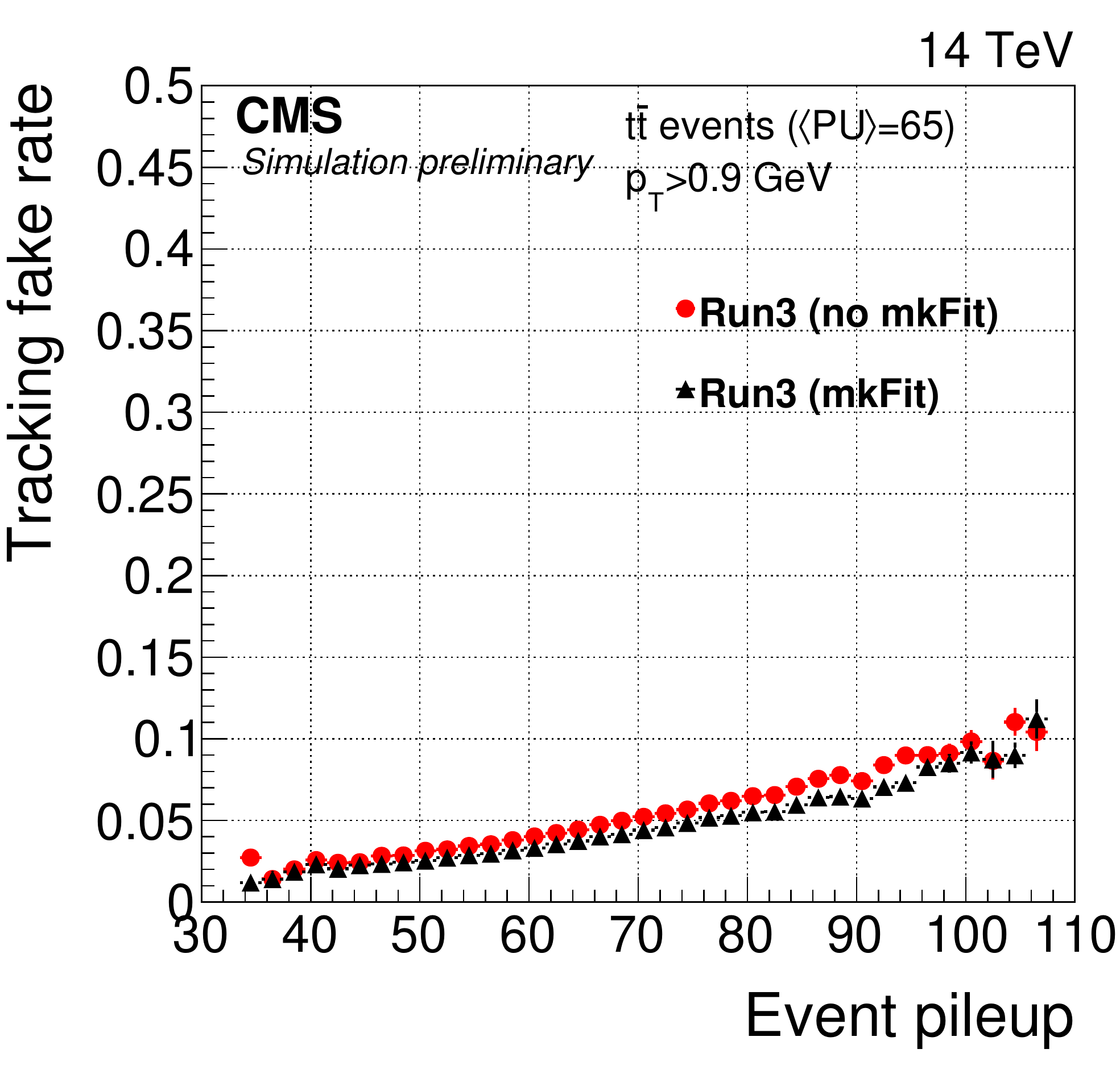}
        \vspace{-4ex}
        \caption{\label{fig:fakerate_pu}Tracking fake rate as a function of the event pileup for CKF tracking (red) and \mkFit\ tracking (black), for reconstructed tracks with $\Pt > 0.9 \GeV$~\cite{mkFitDP}.}
    \end{minipage}
\end{center}
\end{figure}

The \mkFit\ algorithm is used by a subset of the CMS tracking iterations that reconstruct almost 90\% of the hard scattering event tracks with $\Pt > 0.5\GeV$. The physics performance achieved by the usage of the \mkFit\ algorithm for the reconstruction of tracks, identified by the ``high purity" quality flag~\cite{CMSTrackingPaper} and measured in a \ttbar sample with event pileup (PU), i.e. simultaneous collisions, following a Poisson distribution with a mean value uniformly distributed between 55 and 75, is illustrated in figures~\ref{fig:efficiency_pt}-\ref{fig:fakerate_pu}. Starting from figures~\ref{fig:efficiency_pt} and \ref{fig:efficiency_eta}, these show the tracking efficiency, defined as the fraction of simulated tracks matched to at least one reconstructed track, where the matching requires 75\% common hits between the simulated and the reconstructed track, as a function of \Pt and $\eta$ respectively. The efficiency with and without the \mkFit\ algorithm for track building is comparable, with small gains in the pseudorapidity range $2.4 < |\eta| < 2.8$ for the \mkFit\ case. In figures~\ref{fig:fakerate_eta} and \ref{fig:fakerate_pu}, the tracking fake rate, i.e. the fraction of misidentified reconstructed tracks, can be seen as a function of $\eta$ and event PU. When the \mkFit\ algorithm is used, the fake rate tends to be lower for increasing $\eta$ and the scaling with PU is better. Finally, the tracking duplicate rate, which is defined as the fraction of reconstructed tracks associated multiple times to the same simulated track, is marginally increased ($\sim\!0.5\%$) for the \mkFit\ track reconstruction, due to the parallel nature of the algorithm~\cite{mkFitDP}. Duplicate tracks produced by the \mkFit\ algorithm are mitigated by a dedicated duplicate removal, tuned as a function of \Pt and $\eta$.

\begin{figure}[tbh!]
\begin{center}
    \begin{minipage}{0.45\textwidth}
        \includegraphics[width=0.95\textwidth]{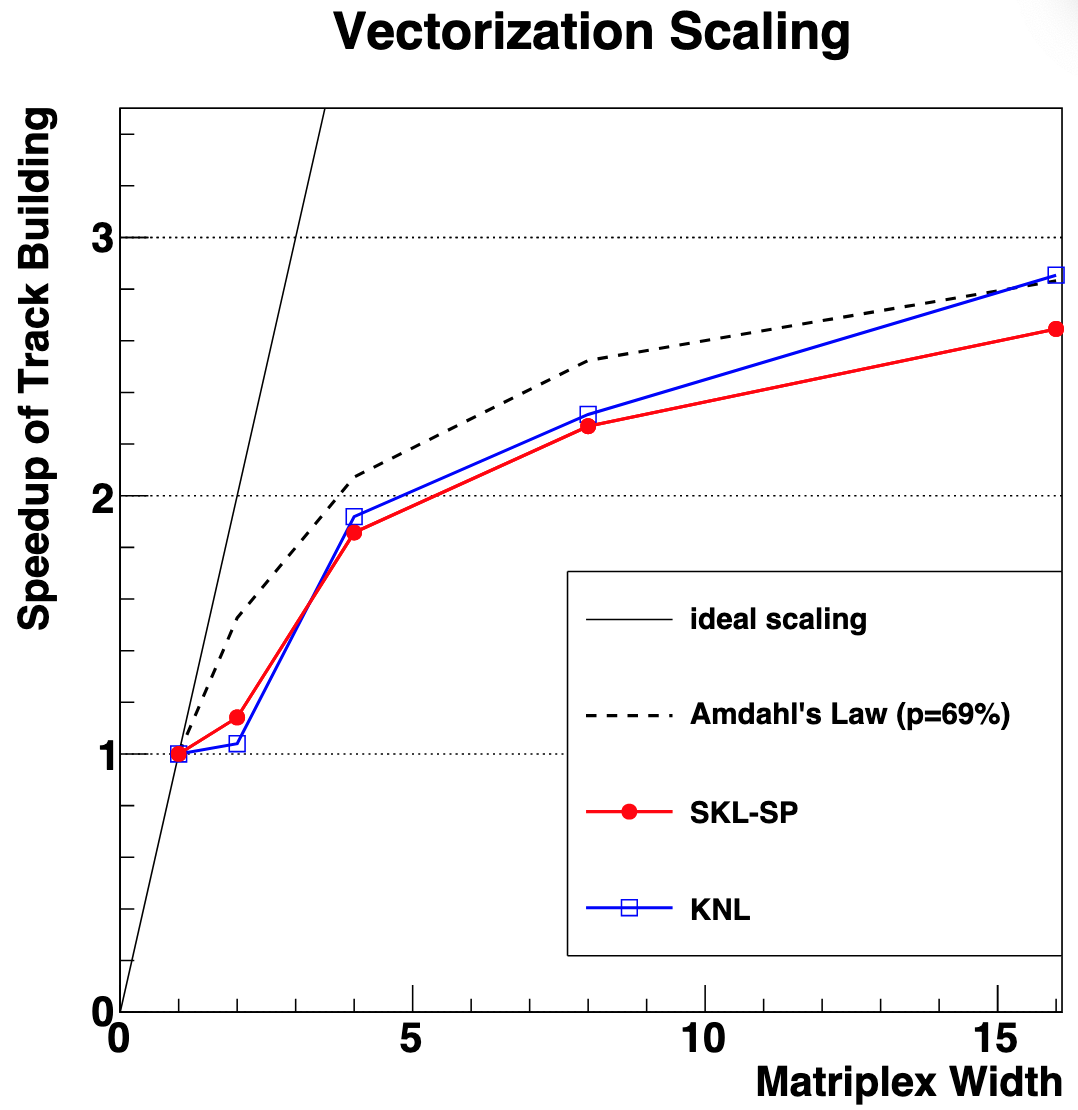}
        \vspace{-1ex}
        \caption{\label{fig:VectorizationScaling}Speedup due to vectorization as a function of the \mtplx\ width for \mkFit\ track building on the KNL and SKL-SP machines. The ideal speedup (solid line) and the speedup based on Amdahl’s Law (dashed lines) are also shown~\cite{mkFitPaper}.}
    \end{minipage}
    \hspace{2pc}
    \begin{minipage}{0.43\textwidth}
        \includegraphics[width=\textwidth]{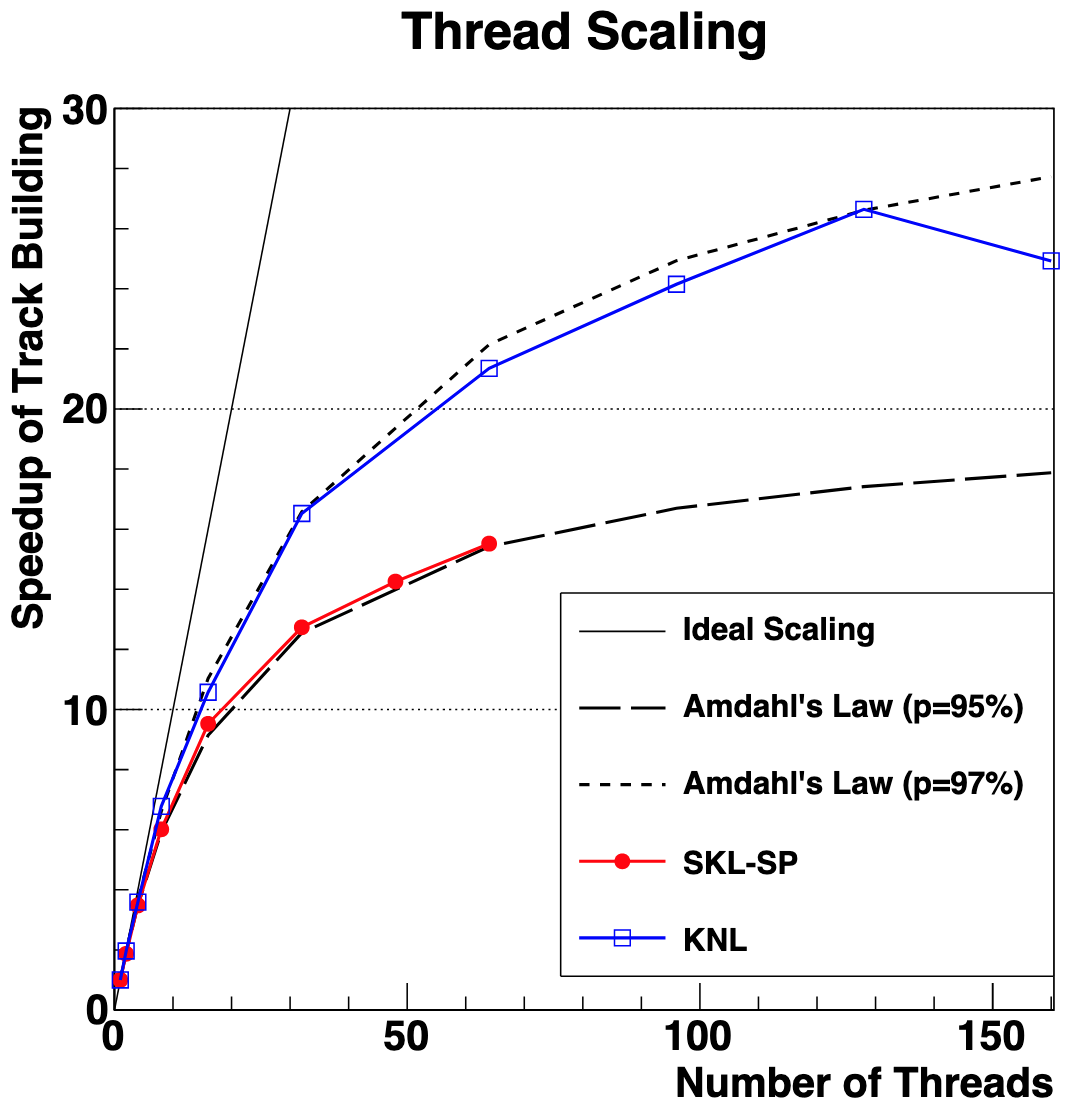}
        \vspace{-4ex}
        \caption{\label{fig:ThreadScaling}Speedup due to multithreading as a function of the number of threads for \mkFit\ track building on the KNL and SKL-SP machines. The ideal speedup (solid line) and the speedup based on Amdahl’s Law (dashed lines) are also shown~\cite{mkFitPaper}.}
    \end{minipage}
\end{center}
\end{figure}

\begin{figure}[tbh!]
\begin{center}
    \begin{minipage}{0.45\textwidth}
        \includegraphics[width=0.95\textwidth]{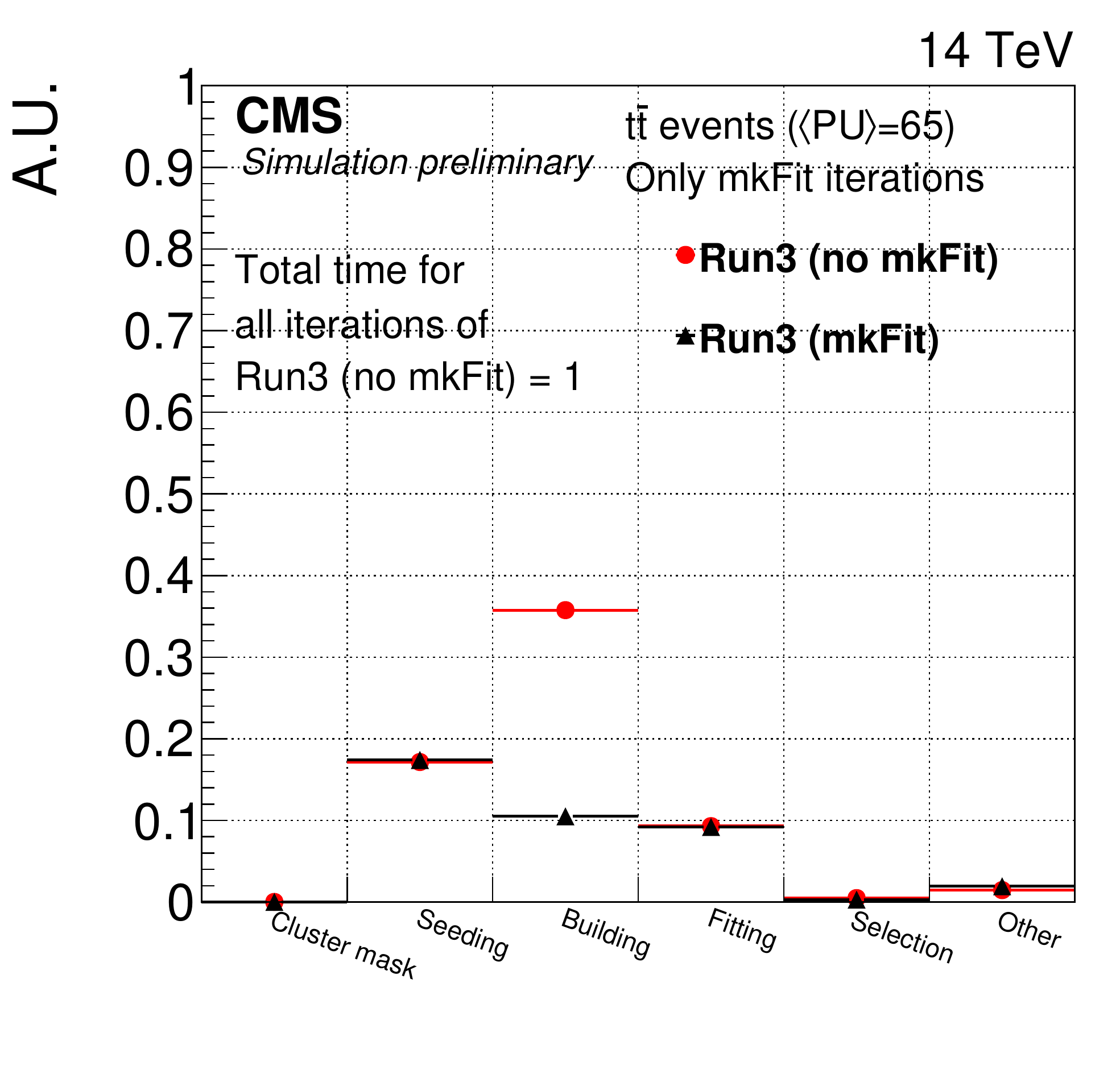}
        \vspace{-4ex}
        \caption{\label{fig:MkFitSteps_realtime}Tracking time as a function of the tracking steps for CKF tracking (red) and \mkFit\ tracking (black), for the subset of tracking iterations using the \mkFit\ algorithm. The vertical axis is normalized to have the total time without \mkFit\ equal to unity.~\cite{mkFitDP}.}
    \end{minipage}
    \hspace{2pc}
    \begin{minipage}{0.43\textwidth}
        \vspace{-3ex}
        \includegraphics[width=\textwidth]{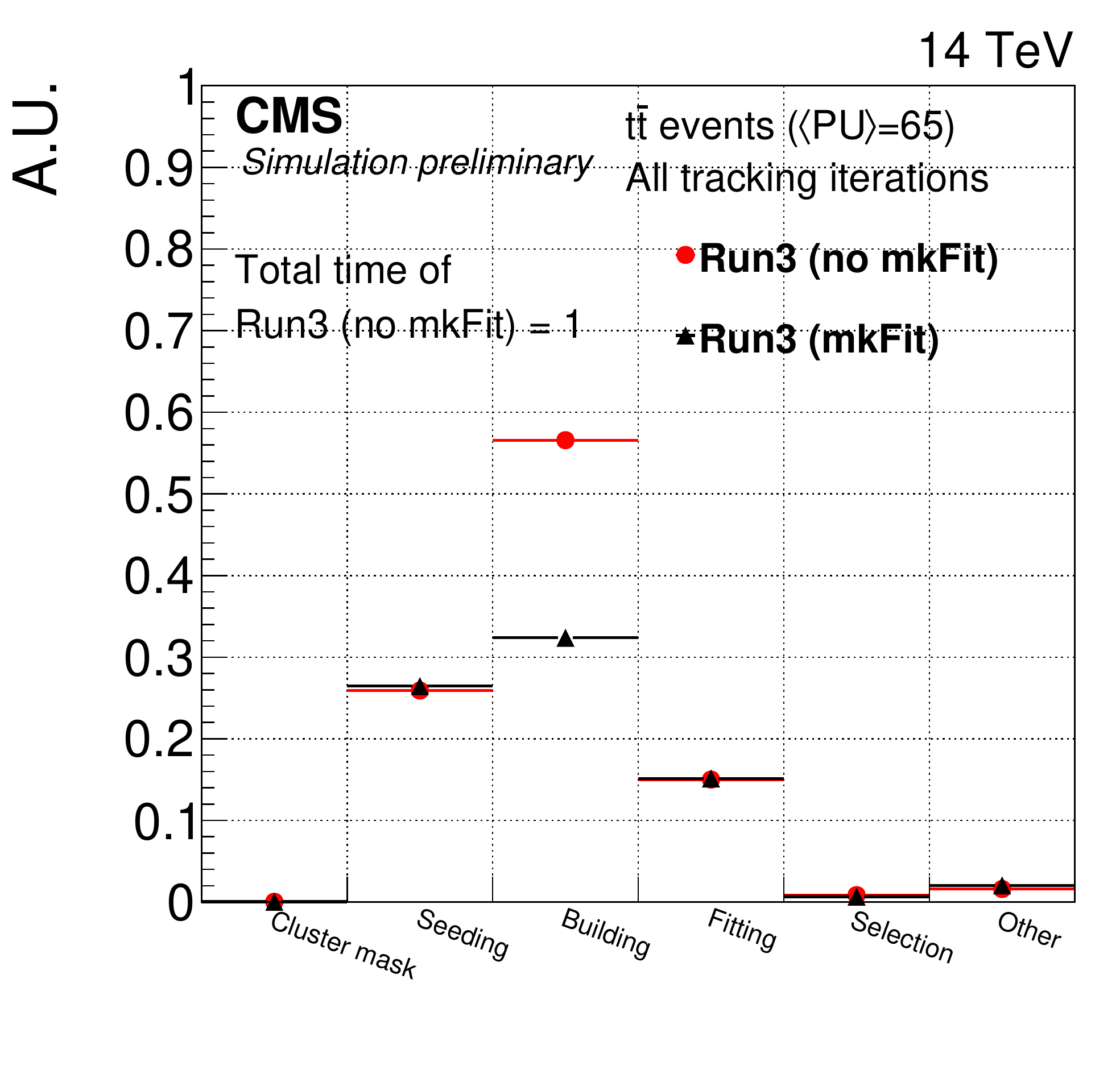}
        \vspace{-7ex}
        \caption{\label{fig:realtime}Tracking time as a function of the tracking steps for CKF tracking (red) and \mkFit\ tracking (black), for all tracking iterations. The vertical axis is normalized to have the total time without \mkFit\ equal to unity.~\cite{mkFitDP}.}
    \end{minipage}
\end{center}
\end{figure}

The great advantage of the \mkFit\ algorithm is evident when the timing performance is measured. In figure~\ref{fig:VectorizationScaling}, the speedup achieved by vectorizing the code is shown as a function of the vector unit width of the \mtplx\ library, i.e. the number of matrices operated on concurrently, while figure~\ref{fig:ThreadScaling} shows the speedup coming from making the code multithreaded as function of the number of threads. The measurements were performed in such a way that the vectorization and the multithreading effects were factorized and on two separate machines: the ``KNL" machine (64 cores: Intel$^\circledR$ Xeon Phi\texttrademark{} processor 7210 @ 1.30 GHz) and the ``SKL-SP" machine (dual socket $\times$ 16 cores: Intel$^\circledR$ Xeon$^\circledR$ Gold 6130 processor @ 2.10 GHz). A comparison with the Amdahl’s Law indicates that almost $70\%$ of the \mkFit\ operations are effectively vectorized and more than $95\%$ are effectively parallelized. The timing performance within the context of the CMS track reconstruction is shown in figures~\ref{fig:MkFitSteps_realtime} and \ref{fig:realtime}. These indicate that, when using the \mkFit\ algorithm, individual \mkFit\ iterations get a building time reduction up to $\times6.7$ and the sum of \mkFit\ iterations get a $\times3.5$ building time reduction (figure~\ref{fig:MkFitSteps_realtime}), while the sum of all iterations gets approximately $\times1.7$ speedup (figure~\ref{fig:realtime}). As a result, the total Run-3 tracking time is reduced by 25\%, which translates in an increase of event throughput of 10--15\%~\cite{mkFitDP}.

\section{Summary and Outlook} \label{sec:SummaryAndOutlook}

\mkFit\ is a Kalman-filter algorithm for track pattern recognition, successfully following the paradigm shift towards code vectorization and parallelization. It has recently been integrated to the CMS central software, replacing the CKF-based tracking in a subset of the CMS tracking iterations for Run-3. The usage of the \mkFit\ algorithm leads to significant improvements in terms of computational performance, while retaining a comparable physics performance.

Looking to the future, more \mkFit-related developments are foreseen to further enhance the CMS tracking. Ideas for further application of the \mkFit\ algorithm include its extension to more track building iterations, its implementation to the track fitting procedure, since it has now become almost as time-consuming as the track building procedure (figures~\ref{fig:MkFitSteps_realtime} and \ref{fig:realtime}), and its application to the High Level Trigger reconstruction of CMS. Finally, the \mkFit\ algorithm is being modified to accommodate the Phase-2 CMS geometry and configuration, while also exploring synergies with other tracking algorithms developed for Phase-2.

\section*{Acknowledgements} \label{sec:Acknowledgements}

This work was supported by the U.S. National Science Foundation under Cooperative Agreements OAC-1836650 and PHY-2121686 and grant NSF-PHY-1912813.

\section*{References}
\bibliographystyle{iopart-num}
\bibliography{mkFit_ACAT2022Proceedings}

\end{document}